\documentclass[pre,twocolumn,floatfix,aps]{revtex4}
\usepackage{amsmath}
\usepackage{makeidx}
\usepackage{amssymb}
\usepackage{graphicx}

\usepackage[usenames]{color}
\usepackage[normalem]{ulem}

\usepackage{hyperref}
\usepackage[T1]{fontenc} 

\begin{document}

\title{Quasi-two-dimensional soliton in a self-repulsive spin-orbit-coupled dipolar binary condensate }

\author{S. K. Adhikari\footnote{sk.adhikari@unesp.br      \\  https://professores.ift.unesp.br/sk.adhikari/ }}

\affiliation{Instituto de F\'{\i}sica Te\'orica, Universidade Estadual Paulista - UNESP, 01.140-070 S\~ao Paulo, S\~ao Paulo, Brazil}
      
%%%%%%%%%%%%%%%%%%%%%%%%%%%%%%%%%%%%%%%%%%%%%%%%%%%%%%%%%%%%%%%%%%%%%%%%%%%%%%
%%%%%%%%%%                    Abstract                             %%%%%%%%%%%
%%%%%%%%%%%%%%%%%%%%%%%%%%%%%%%%%%%%%%%%%%%%%%%%%%%%%%%%%%%%%%%%%%%%%%%%%%%%%%

\date{\today}

\begin{abstract}

We study  the formation of  solitons 
in a  {\it  uniform}   quasi-two-dimensional (quasi-2D)   spin-orbit (SO) coupled self-repulsive binary  dipolar   and nondipolar
Bose-Einstein condensate (BEC)   using  the mean-field Gross-Pitaevskii equation. 
For a weak SO coupling, in a nondipolar BEC,  one  can have three types of degenerate solitons:  a  
  multi-ring soliton  with intrinsic vorticity of angular momentum projection $+1$ or $-1$ in one component and 0 in the other,
    a circularly-asymmetric soliton  and  a stripe soliton with stripes in the density.  
For an intermediate  SO couplings,  the multi-ring soliton ceases to exist and there appears a square-lattice soliton with a spatially-periodic pattern in density on a square lattice, in addition to the degenerate  circularly-asymmetric and stripe solitons. 
 In the presence of a dipolar interaction, with the polarization direction aligned in the quasi-2D plane, only     the degenerate  circularly-asymmetric and stripe solitons appear.  

\end{abstract}

 \maketitle

\section{Introduction}

In mathematics and physics, a soliton is a nonlinear,  one-dimensional  (1D) self-reinforcing, localized wave packet that is strongly stable due
to a balanced cancellation of attractive nonlinear and repulsive dispersive effects in the medium. 
 A soliton can be found in diverse systems, for example, in a 
  water wave,  in a nonlinear medium \cite{nlm}, in  nonlinear optics \cite{x2} and  in a   Bose-Einstein condensate (BEC) \cite{x1}.  Of these,  a soliton in a BEC is especially interesting as it is 
 a quantum  system. In  a quasi-one-dimensional
  (quasi-1D) set up, solitons have been created and studied in a  BEC
of $^7$Li \cite{r2a,r2b} and $^{85}$Rb \cite{r3} atoms, following the suggestion of  a theoretical investigation  \cite{r5}, 
using  the mean-field Gross-Pitaevskii (GP) equation \cite{r6a,r6b}. 
 There are numerous examples
of similar quasi-1D solitons in BECs \cite{38,39,40,41,42,43,44,45}.
Solitons have also been studied in binary BEC mixtures \cite{r7a,r7b}. 
  However, stable solitons cannot be formed  in two (2D) \cite{townes} and three dimensions (3D) \cite{x1,x2}  due to a collapse instability.

Soon after the observation  \cite{BEC} of a BEC of $^{87}$Rb \cite{rb},  $^7$Li \cite{li} and  $^{23}$Na \cite{na} atoms in a laboratory,  it was possible to observe and study  a hyperfine
spin-one multi-component spinor BEC of $^{23}$Na atoms \cite{exptspinor}. Although it is not possible to have  a natural spin-orbit (SO) coupling in  electrically neutral BEC atoms,  an artificial synthetic SO coupling  can be created in neutral atoms by  tuning  a few
Raman laser  beams, which couple  the different hyperfine spin states
\cite{thso,exptsob,sodip2}. Dresselhaus \cite{SOdre} and  Rashba \cite{SOras}  suggested 
two possible ways of creating such SO couplings. 
An equal mixture of  
Dresselhaus  and Rashba SO couplings has been experimentally realized 
in a   BEC of $^{87}$Rb  \cite{exptso1}
and  $^{23}$Na \cite{na-solid} atoms 
 of  $F_z=0,-1$ hyperspin components.
The mean-field binary GP equation of such an SO-coupled    BEC  is coupled by the Pauli spin matrices and hence this BEC is called  a pseudo spin-half BEC.  
  Multi-component spinor  BECs   possess distinct properties and can have different 
  excitation  \cite{kita1,kita2}, which are  not allowed  in a single-component scalar BEC.       For example, a quasi-1D \cite{quasi-1d}, a quasi-two-dimensional (quasi-2D) \cite{quasi-2da,quasi-2db} or a   3D  \cite{quasi-3d}  soliton   can be stabilized  in 
an SO-coupled binary BEC, which can have distinct spatial structure. 
%Possible ways of realizing the SO coupling in a three-component   spin-one BEC have
%been considered \cite{3c}. 
% which is a simplification over the three-component spin-one hyperfine spin state 5S$_{1/2}$ of $^{87}$Rb.   
 Later,  SO-coupled BECs were   studied   \cite{exptsob} and 
an SO-coupled 
 spin-one condensate of $^{87}$Rb atoms  in  hyperspin states  $F_z =\pm 1,0$ was  investigated  \cite{exptsp1}.
%A three-component SO-coupled spin-one BEC is  known to exhibit a rich variety of physical phenomena \cite{thspinor}
%not possible in a two-component pseudo spin-half BEC.  
A   {spin-one} SO-coupled multi-component BEC also 
sustains a quasi-1D \cite{quasi-1d1} or a quasi-2D \cite{quasi-2d1} or a 3D 
\cite{quasi-3d1} soliton,  not possible in a single-component BEC \cite{townes}.  
% Stripe and other structures in density have been found in a Rashba coupled pseudo spin-half and spin-one 
%BEC \cite{referee1}.

 The studies of quasi-2D \cite{quasi-2d1} and 3D  \cite{quasi-3d1} solitons in an SO-coupled  nondipolar binary BEC indicate that the SO coupling introduces a stabilizing attractive interaction necessary for the formation of a self-bound soliton.  {  All previous studies on SO-coupled nondipolar  BEC solitons were performed on a self-attractive
  system with a negative atomic scattering length \cite{quasi-2da,quasi-2db,quasi-2d1}  corresponding to an attraction among the atoms, which facilitates the formation of a soliton.  A positive atomic scattering length corresponds to a repulsion among the atoms \cite{BEC}.
  However, the scattering length in a BEC can often be varied continuously from positive to negative, thus making the system attractive from repulsive, by changing an 
external background magnetic field near a Feshbach resonance \cite{fesh,fesh1}. In fact, this technique was used to change the scattering length  of  $^{85}$Rb atoms in the BEC from a positive value   to a negative value to obtain   a  self-attractive BEC  of    $^{85}$Rb atoms   and a quasi-1D soliton was created and studied \cite{r3} in this system  in a laboratory.  }
  In a different front it was demonstrated  \cite{malomed} that it is  possible to have a  stable anisotropic soliton in a quasi-2D dipolar BEC with the polarization direction lying in the quasi-2D plane.   There have also been many studies of quasi-1D
solitons in dipolar BECs under different conditions \cite{30,31,32,33}.
  
  { The long-range nonlocal anisotropic  dipolar interaction sets up a different scenario  \cite{malomed} for the stabilization of a quasi-2D soliton, which will not be possible in the absence of a dipolar interaction \cite{townes}. Many new phenomena are possible in a dipolar BEC, such as, angular collapse \cite{bohn},  roton mode \cite{roton2,roton1},
 quasi-2D asymmetric soliton \cite{malomed},  quantum droplet  \cite{drop}, supersolidity in 1D \cite{ss1d} and 2D \cite{ss2d}, anisotropic expansion \cite{aaexp} etc.   The study of a dipolar BEC got new impetus after the experimental observation of dipolar BECs  of $^{52}$Cr \cite{1,2,6}, $^{166}$Er \cite{7}, $^{168}$Er \cite{8}, and  $^{164}$Dy
\cite{9,11}  atoms, with large magnetic
dipole moment.

  In contrast to previous studies \cite{kono,adhisol,quasi-2da,quasi-2db} of SO-coupled  binary quasi-2D solitons in a self-attractive BEC,
here we perform  a comprehensive study of quasi-2D solitons in a binary
 Rashba or a Dresselhaus SO-coupled self-repulsive  nondipolar and also dipolar  BEC with positive scattering lengths, where  the polarization direction of the dipolar atoms lies in the quasi-2D plane.
  The solitons in this system are bound due to a joint action of the SO coupling and the long-range dipolar interaction, in spite of the short-range contact repulsion in a self-repulsive binary BEC and new types of 2D solitons are expected in this system.  We find drastic changes in soliton structure and density in an SO-coupled binary BEC soliton after the introduction of a weak dipolar interaction.}   There have been a few previous studies on  different aspects of an   SO-coupled dipolar BEC \cite{sodip1,sodip2,sodip3,sodip4,sodip5,sodip6,sodip7,yuka,yuka1}.

 We find  different types  solitons in an SO-coupled quasi-2D binary nondipolar self-repulsive  BEC.  For a small value of SO-coupling strength parameter $\gamma$ ($0 < \gamma \lessapprox 1$) there can be three types of degenerate solitons.
   The first type is a  $(\mp 1,0)$-type multi-ring soliton,  where the numbers in the  parenthesis represent  the  angular momentum projections of  the two components with the upper (lower) sign corresponding to Rashba (Dresselhaus) SO coupling. This  state has  1/2 unit of  angular momentum projection  and  is often called  a half-quantum vortex state \cite{hpu}. This half-quantum vortex state has  been experimentally observed and studied in a topological superfluid $^3$He \cite{half-vortex,hv}.  The second type is a  stripe soliton with  a spatially-periodic stripe pattern in density of the two components and no  such stripe in the total density.   Such stripe pattern in a pseudo spin-half SO-coupled  BEC of
  $^{23}$Na  atoms was experimentally observed \cite{na-solid}.
 The third type is a   circularly-asymmetric soliton.  
   {The component densities and total density of a  circularly-asymmetric soliton are  also  circularly-asymmetric.} 
   For medium SO coupling ($\gamma \gtrapprox 1$),  
the aforementioned half-quantum vortex multi-ring soliton ceases to exist and  
 we find three different types of degenerate solitons  in a nondipolar binary BEC, e.g.,   a    stripe soliton,  a  circularly-asymmetric soliton, and  a square-lattice soliton  with a spatially periodic 2D structure along both $x$ and $y$ directions.
 % (d) a four-wing soliton  with four wings along $x$ and $y$ axes.
  
  The aforementioned spatially-periodic stripe \cite{14,baym1,baym2,2020,stripe}
and square-lattice \cite{adhisol} pattern in an SO-coupled BEC is
intimately related to a similar pattern in a supersolid \cite{s1,s2,s3,s4,s5}. A supersolid is a special quantum state of matter
with a rigid spatially-periodic crystalline structure \cite{54},
breaking continuous translational invariance, that flows
with zero viscosity as a superfluid breaking continuous
gauge invariance.

The introduction of a dipolar interaction in an SO-coupled BEC changes the scenario of soliton formation \cite{malomed}.  
The dipolar interaction, with the polarization direction in the quasi-2D $x$-$y$ plane,  breaks the circular symmetry in this plane.  Consequently,  the quasi-2D soliton  gets stretched along the polarization direction. Because of this stretching, 
the square-lattice soliton %and the four-wing soliton 
ceases to exist and in this case we have only  two types of degenerate solitons in a quasi-2D self-attractive or self-repulsive
SO-coupled dipolar binary BEC:  stripe solitons and  circularly-asymmetric solitons.  
For all cases studied in this paper,  the density  and energy of the  
solitons are the same for Rashba and Dresselhaus SO couplings, although  the two wave functions are different.   In the presence of an SO coupling, magnetization is not conserved. However, the magnetization of most solitons studied in  this  paper is zero. 
   
 In Sec. \ref{i} we present the mean-field GP equation for the SO-coupled BEC in dimensionless form appropriate for the study of a quasi-2D soliton. In Sec. \ref{ii} we present an analytic study of the eigenfunctions  of the quasi-2D  solitons  and identify the origin of the degenerate multi-ring, stripe, and square-lattice solitons. In Sec. \ref{iii}  we present the numerical results of this investigation. In the nondipolar SO-coupled self-repulsive and self-attractive quasi-2D BEC   we identify four types of solitons as elaborated in Sec. \ref{a}: the multi-ring,  stripe, circularly-asymmetric, and square-lattice  solitons.  However, in the dipolar self-repulsive and self-attractive  cases we find   
only two types of solitons as presented in Sec. \ref{b}: the   stripe and circularly-asymmetric solitons. Finally, in Sec.  \ref{iv} we present a brief summary of this investigation.  

\section{Mean-field model}
\subsection{Gross-Pitaevskii equation}

\label{i}

%For the study of soliton formation in  a  quasi-2D  spin-one spinor BEC, 
We consider a binary SO-coupled   BEC of $N$ atoms, two components $j=1,2$, each of mass { $m$}, under a
harmonic trap $V({\bf r})= m\omega_\rho^2(x^2+y^2)/2+ { m}\omega_z^2 z^2/2$ $(\omega_z \gg \omega_\rho)$ of  frequency $\omega_z$ along the $z$ axis and 
and $\omega_\rho$     in the  quasi-2D $x$-$y$ plane. The single particle Hamiltonian of the SO-coupled  BEC is
%
%The single-particle Hamiltonian 
%of the condensate without atomic interaction   in this quasi-2D trap in dimensionless  variables
 \cite{exptso1} 
\begin{align}\label{sp}
H_0 =-\frac { \hbar^2}{2 m}(\partial_x^2+\partial_y^2+\partial_z^2)   +  V({\bold r})+\gamma [\eta p_y \sigma_x -  p_x \sigma_y],
\end{align}
%\begin{equation}
%H_0 = \frac{p_x^2+p_y^2}{2{  m}}  +H_{\mathrm{SO}}  ,
%\label{sph} 
%\end{equation}
where ${\bold r}\equiv \{x,y,z   \}$,  ${\boldsymbol \rho} \equiv \{ x,y\}$,
  $\partial_x = \partial/\partial x$ etc.,
$\gamma$ is the strength of the SO-coupling interaction,
%where %${\bf r}=\{x,y,z\}$, $\nabla^2_{\bf r}=p_x^2+p_y^2+p_z^2,$
%$p_x = -i\hbar \partial_x \equiv -i\hbar \partial/\partial x, p_y = -i\hbar \partial_y \equiv -i\hbar %\partial/\partial y $. 
%are the momentum operators along $x$ and $y$ axes, respectively,  .
{
for Rashba (Dresselhaus) coupling the parameter $\eta=+1$ ($\eta=-1$), 
%{ and $\eta=0$ for an equal mixture of 
%Rashba and Dresselhaus couplings,}
%{multi For Rashba and Dresselhaus SO couplings, respectively, the SO-coupling term } \cite{exptso1}     nodip-gma-2-rep-str-FgB
%\begin{align}
%$H_{\mathrm{SO}}= - \gamma p_x \sigma_y+\gamma p_y \sigma_x,$ and
%$$H_{\mathrm{SO}}=  - \gamma p_x \sigma_y-\gamma p_y \sigma_x,$
%\end{align}
  $p_x=-{\mbox i}\hbar \partial_x$ and  $ p_y=-{\mbox i}\hbar \partial_y$, with ${\mbox i} =\sqrt{-1},$ are the momentum along $x$ and $y$ axes, respectively, and }
the Pauli spin matrices  $\sigma_x$ and $\sigma_y$ are
\begin{eqnarray}
\sigma_x=\begin{pmatrix}
0 & 1  \\
1 & 0 
\end{pmatrix}, \quad  \sigma_y=\begin{pmatrix}
0 & -{\mbox i}\\
{\mbox i} & 0  
\end{pmatrix}.
\end{eqnarray}
%and $\gamma$ is the strength of SO coupling.  
%The numerical results for density and energy of the solitons are the same for both types of SO coupling, although the respective wave functions are different.

The interaction between the two atoms has the following two parts: the contact interaction $\delta({\bf R})$ and the nonlocal asymmetric dipolar interaction $V_{\mathrm{dd}}({\bf R}).$
The  intraspecies ($V_j, j =1,2$) and 
interspecies ($V_{12}$)
interactions 
for two atoms  at positions $\bf r$ and $\bf r'$ are given by \cite{1}%  \cite{crrev,expt}
\begin{eqnarray}\label{intrapot} 
V_j({\bf R})&=&
\frac{\mu_0\mu^2}{4\pi}V_{\mathrm{dd}}({\mathbf R})+\frac{4\pi 
\hbar^2 a_j}{m}\delta({\bf R }),
 \\
 \label{interpot} 
V_{12}({\bf R})&=& \frac{\mu_0\mu^2}{4\pi}V_{\mathrm{dd}}({\mathbf R})+
\frac{4\pi \hbar^2 a_{12}}{m}\delta({\bf R}),\\ \label{Y. Tang1,2, A. G. Sykes3, N. Q. Burdick2,4, J. M. DiSciacca2,4, D. S. Petrov3, and B. L. Lev1,dp}
V_{\mathrm{dd}}({\mathbf R})  &=& 
\frac{1-3\cos^2\theta}{{\mathbf R}^3},
     \end{eqnarray}
where $\bf R \equiv (r-r')$ is the position vector joining the two atoms, $\mu_0$ is the permeability of free space, $\mu$  is the magnetic dipole moment of each atom,
$\theta$ is the angle made by the vector ${\bf R}$ with the polarization 
$y$ direction,  $a_j$ is the intraspecies scattering length, and $a_{12}$  is the interspecies  scattering length. The scattering length measures the strength of contact interaction in an ultra-dilute BEC. 
To compare the dipolar and contact interactions, the  
dipolar interaction  is  expressed in terms of the dipolar length, defined by \cite{1}
\begin{align}
  a_{\mathrm{dd}}\equiv 
\frac{\mu_0\mu^2m}{12\pi \hbar ^2}.  \quad  
\end{align}

{ In the actual experimental realization of an SO-coupled  binary dipolar BEC of equal-mass atoms we can consider a binary $^{164}$Dy-$^{162}$Dy BEC. The atomic masses of these two isotopes of the dysprosium atom are almost equal, which justifies the equal mass approximation.  The magnetic dipole moment of each species is $\mu=10\mu_B$, where $\mu_B$ is a Bohr magneton.
   Hence the two species of atoms have the  same dipole moment, same  intraspecies and  interspecies dipolar interactions and the   dipolar lengths of the two components are  equal. As the nuclear spin of both atoms $-$   $^{164}$Dy and $^{162}$Dy  $-$ is zero, there is no hyperfine structure.  Nevertheless, a
synthetic SO coupling can be established in this binary system involving different types of atoms, particularly in the context of ultracold atomic gases used for quantum simulation. 
 Dipolar BECs of  both $^{164}$Dy \cite{9,pq2} and $^{162}$Dy \cite{pq3,pq4} atoms have been experimentally realized 
 \cite{1x} and are now routinely studied in different laboratories.  The interspecies and intraspecies scattering lengths of this binary dipolar BEC are all positive \cite{lq1,1x},  corresponding to a self-repulsion among the ingredients, which will make the formation of a soliton difficult.

 In the case of a nondipolar SO-coupled binary mixture we consider two hyperfine spin states, 
$F_z=0,-1$,
of the $^{87}$Rb atom with positve  scattering lengths \cite{lq3}. Such an SO-coupled mixture has already been observed \cite{exptso1} and studied in a laboratory \cite{exptso2}.  
This binary BEC seems to be an ideal system for the study of a self-repulsive SO-coupled binary mixture with equal atomic masses. 
}

The dipolar  binary BEC  of two components $j=1,2$  is described by the following set of equations \cite{1,bao}
\begin{align}
%\begin{eqnarray}
\label{Eq1}
{\mbox i} \hbar \partial_t \phi_j({\bf r},t) &=\frac{\hbar^2}{m}
\left[  -\frac{1}{2}(\partial_x^2+\partial_y^2+\partial_z^2)+ \frac{m^2}{2\hbar^2}  
(\omega_\rho ^2\rho^2+\omega_z^2 z^2 )  \right.
%\end{eqnarray}
\nonumber
\\
%\begin{eqnarray}
&+ 4\pi {m}{a}_j N_j \vert \phi_j({\bf r},t) \vert^2
+4\pi  {a}_{12} N_k \vert \phi_k({\bf r},t) \vert^2 \nonumber\\
%\end{eqnarray}
%\\ &
%\begin{eqnarray}
%\nonumber 
&+{3}a_{\mathrm{dd}} N_j \int  V_{\mathrm{dd}} ({\mathbf R})\vert\phi_j({\mathbf r'},t)\vert^2 d{\mathbf r}' \nonumber \\
&+3a_{\mathrm{dd}} N_k \left. \int  V_{\mathrm{dd}} ({\mathbf R})\vert\phi_k
({\mathbf r'},t)\vert^2 d{\mathbf r}' 
\right] 
 \phi_j({\bf r},t),%\end{align}
 \nonumber \\
 &j \ne k=1,2,
%\end{eqnarray}
\end{align}
where   $\partial_t \equiv \partial/\partial t$,  $N_j$ and $N_k$ are the number of atoms in the two components ($N=N_1+N_2$).  The dipolar atoms  are polarized along the $y$ axis.
Equations (\ref{Eq1}) can be cast in the following dimensionless form if we scale lengths in units of $l_0=\sqrt{\hbar/m\omega_z}$, time units of $t_0=1/\omega_z$,  
density  in units of $l_0^{-2}$, and  energy in units of $\hbar \omega_z$ 
\begin{align}
{\mbox i}  \partial_t \phi_j({\bf r},t)&=
\left[  -\frac{1}{2}(\partial_x^2+\partial_y^2+\partial_z^2)+ \frac{1}{2}  
\left(\frac{\omega_\rho ^2}{\omega_z^2}\rho^2+ z^2 \right) \right.
%\end{align}
\nonumber\\
%\begin{align}
 &
+ {4\pi }{a}_j N_j \vert \phi_j({\bf r},t) \vert^2
+{4\pi } {a}_{12} N_k \vert \phi_k({\bf r},t) \vert^2
%\end{align}
\nonumber\\
%\begin{align}
&+{3}a_{\mathrm{dd}} N_j \int  V_{\mathrm{dd}} ({\mathbf R})\vert\phi_j({\mathbf r'},t)\vert^2 d{\mathbf r}' \nonumber
 \\
&+\left. {3}a_{\mathrm{dd}} N_k \int  V_{\mathrm{dd}} ({\mathbf R})\vert\phi_k({\mathbf r'},t)\vert^2 d{\mathbf r}'
\right] \phi_j({\bf r},t), %\nonumber \\
%&\times  \phi_j({\bf r},t),\quad  \quad j\ne k=1,2.
\label{eqp}
\end{align}
where $ j\ne k=1,2.$
Without any risk of confusion,
here and in the following we are using the same symbols to denote the scaled dimensionless and unscaled variables.

We consider a quasi-2D  binary dipolar BEC with a strong trap along the $z$ direction.  We assume that the dynamics of the BEC in the $z$ direction  is frozen in the ground state 
\begin{equation}
\psi_B(z) ={\pi}^{-1/4} e^{-z^2 /2}.
\end{equation}
and the relevant dynamics of the dipolar BEC will be confined in the $x$-$y$ plane. In this case it is possible to integrate out the $z$ variable and write a set of binary  coupled equations for the relevant dynamics in the $x$-$y$ plane. 
The component wave function of the BEC can be written as 
\begin{equation}\label{pqr}
\phi_j({\bf r},t)=\psi_B(z)\times \psi_j({\boldsymbol \rho},t),
\end{equation}
where $\psi_i({\boldsymbol \rho},t)$ is the wave function of the quasi-2D BEC confined in the $x$-$y$ plane.
To obtain the relevant dynamics, we substitute  Eq. (\ref{pqr}) into Eq. (\ref{eqp}), multiply the resultant equation by $ \psi_B(z)$ and integrate over $z$ \cite{2d-3d,bec2015}. To avoid the difficulty associated with the integration over the divergent dipolar interaction in the configuration space, that integral was evaluated in the momentum space.  
This procedure     \cite{2d-3d} leads to the   
 following set of  GP equations   for the two  components  \cite{bec2015,bao}
\begin{align}
%\begin{eqnarray}
\label{EQ1} 
{\mbox i} \partial_t \psi_{j}({\boldsymbol 
\rho},t)&= \left[ - \frac{1}{2}(\partial_x^2+\partial_y^2)
+c_1 n_j({\boldsymbol \rho},t)  + c_2n_k({\boldsymbol \rho},t)\right. \nonumber %\\
\end{align}
\begin{align}
 &+  \left. d\sum_{j=1,2} s_j({\boldsymbol \rho},t)  \right] \psi_{j}({\boldsymbol \rho},t) 
- { \gamma} [{\mbox i}\eta\partial_y  \nonumber \\
 & + (-1)^j \partial_x]  \psi_{k}  ({\boldsymbol \rho},t)  \, ,\quad \quad  j\ne k=1,2,
\\
%\end{align}
%\\
%{\mbox i} \partial_t \psi_{1}({\boldsymbol 
%\rho},t)&= \Big[ - \frac{1}{2}\nabla^2_{\boldsymbol \rho}
%+{c_1}
%n_{1}({\boldsymbol 
%\rho},t) +c_2 n_2({\boldsymbol 
%\rho},t)   \nonumber \\
% &+d  s_1({\boldsymbol \rho},t) +d  s_2({\boldsymbol \rho},t) \Big]
%\psi_{1}({\boldsymbol 
%\rho},t) \nonumber \\
%&+ { \gamma} (-{\mbox i}\eta\partial_y  + \partial_x)  \psi_{2}  ({\boldsymbol \rho},t)  %%\, , 
%\\
%\label{EQ2}
%i \partial_t\psi_2({\boldsymbol \rho},t)&=\big[ -\frac{1}{2}\nabla^2_{\boldsymbol \rho}+{c_1}
%n_{2} ({\boldsymbol 
%\rho},t)+c_2 n_1({\boldsymbol 
%\rho},t)  \nonumber \\ &+d  s_2({\boldsymbol \rho},t) +d  s_1({\boldsymbol \rho},t) \big] \psi_{2}({\boldsymbol 
%\rho},t)\nonumber \\& - {\gamma} ({\mbox i}\eta\partial_y  + \partial_x)  \psi_{1}  ({\boldsymbol \rho},t)  \, 
% ,\\
 %\end{align}
s_j({\boldsymbol \rho},t) &=
 \int \frac{d{\bf k}_\rho}{(2\pi)^2} e^{-{\mbox i} {\bf k}_\rho\cdot {\boldsymbol \rho}}
\widetilde n_j({\bf k}_\rho,t)j_{2D}\left(\frac{k_\rho }{\sqrt {2 }}  \right) \,  ,
% \nonumber
  \\
%\end{align}
%\begin{align}
 j_{2D}(\xi) &=\frac{1}{2\pi}\int_{-\infty}^{\infty}dk_z \left(\frac{3k_y^2}{{\bf k}^2}-1   \right)|\widetilde n(k_z)|^2, \nonumber \\
   &=\frac{1}{\sqrt{2\pi  }}\left[ -1+3\sqrt \pi \frac{\xi_y^2}{\xi} e^{\xi^2} \left\{ 1- \mathrm{erf}(\xi)\right\} \right]\, , 
 \\
\xi &=\frac{k_\rho}{\sqrt{2}}\, ,\quad \xi_y =\frac{k_y}{\sqrt{2}}\, , \quad k_\rho = \sqrt{k_x^2+k_y^2},%\nonumber %\\
 \\
\widetilde n_j({\bf k}_\rho,t)& =\int d {\boldsymbol \rho} e^{{\mbox i}{\bf k}_\rho \cdot {\boldsymbol \rho}} |\psi_j({\boldsymbol \rho},t)|^2, 
\\
\widetilde n({ k}_z)& =\int_{-\infty}^{\infty}   d {k_z} e^{{\mbox i}{ k}_z z} |\psi_B (z)|^2= e^{-k_z^2/4^2}\, , %\nonumber 
%\end{eqnarray}
\end{align}
where $d=4\pi a_{\mathrm{dd}}N,
c_1 = 2N \sqrt{2\pi}a, c_2=2N \sqrt{2\pi}a_{12}$ and  
we have included the SO-coupling terms  and  assumed that $a_1=a_2\equiv a$.  
Here   
 $n_j({\boldsymbol \rho,t}) = |\psi_j({\boldsymbol \rho,t})|^2, j=1,2$ are the densities of the two components, and $n ({\boldsymbol \rho},t)= \sum_j n_j({\boldsymbol \rho},t)$  the total density. In this study of quasi-2D solitons in the $x$-$y$ plane we have dropped the harmonic oscillator trapping  potential from Eqs. (\ref{EQ1}).
The normalization condition is 
%\begin{align}\label{noma}
$ {\textstyle \int} n({\boldsymbol \rho})\, d{\boldsymbol \rho}=1\, .$
%\end{align}

 Equations  (\ref{EQ1})  can be derived from the energy functional
 \begin{align}
 E[\psi_j] & = \frac{1}{2} \int_{-\infty}^{\infty}  dx\int_{-\infty}^{\infty}  dy \left[  \sum_{j =1}^2 \left( |\partial_x \psi_j|^2 +  |\partial_y \psi_j|^2  \right)
 %|\nabla_{\boldsymbol \rho}\psi_j|^2
 \right. \nonumber \\ & +c_1(n_1^2+n_2^2) +2c_2n_1n_2+ d(s_1^2+s_2^2)+2ds_1s_2
 \nonumber  \\&
   +2\gamma\{\psi_1^*(\partial_x-{\mbox i} \eta\partial_y)\psi_2 -
 \psi_2^* (\partial_x+{\mbox i} \eta\partial_y)\psi_1\}  \Big]\, ,
 \end{align}
 using the variational rule \cite{BEC}
  \begin{align}
  \mathrm{i} \partial_t \psi_j  = \frac{\delta E [\psi_j ]}{\delta \psi_j}
   \end{align}

\subsection{Analytic Consideration} 

\label{ii}

Many  properties of a binary nondipolar or dipolar, self-repulsive or  self-attractive  SO-coupled uniform BEC  can be understood from a consideration  of the eigenfunction-eigenvalue problem of the linear Hamiltonian, setting all nondipolar and dipolar nonlinear interactions and the confining trap  to zero [$c_1=c_2=d=V({\bf r})=0$]. The quasi-2D  stationary wave function of the linear single-particle trapless   Hamiltonian  corresponding to Eq. (\ref{sp})
\begin{equation}\label{abc}
H_{0}^{2D}= -\textstyle \frac{1}{2}  (\partial_x^2+\partial_y^2)
 -{\mbox i}\gamma[\eta \partial_y \sigma_x-\partial_x \sigma_y] 
\end{equation}
 satisfies the following  eigenfunction-eigenvalue problem
\begin{eqnarray}
\begin{bmatrix}\label{spf}
-%\textstyle 
\frac{1}{2}(\partial_x^2+\partial_y^2)&  \gamma (\partial_x-{\mbox i}\eta \partial_y)  \\
-\gamma(\partial_x +{\mbox i}\eta \partial_y) & -%\textstyle
 \frac{1}{2}(\partial_x^2+\partial_y^2)
\end{bmatrix}   \begin{pmatrix} \psi_1 \\ \psi_2 \end{pmatrix}
=  {\cal E}   \begin{pmatrix} \psi_1 \\ \psi_2 \end{pmatrix},
\end{eqnarray}
with energy eigenvalue ${\cal E} $.
In circular coordinates $\boldsymbol \rho =\{\rho, \theta\}, x=\rho\cos \theta, y=\rho\sin\theta$, and we have $(\partial_x\pm {\mbox i}\partial_y) = \exp (\pm {\mbox i}\theta) (\partial_\rho \pm {\mbox i}\partial_\theta /\rho),$ with $\partial_\rho \equiv \partial/\partial\rho, \partial_\theta \equiv \partial/\partial\theta$.  A circularly-symmetric solution  of angular momentum projection $m$ of  Eq.
(\ref{spf}) has the form  \cite{hpu,hv}
\begin{eqnarray}\label{12x} \begin{pmatrix} \psi_1(\boldsymbol \rho) \\ \psi_2 (\boldsymbol \rho)\end{pmatrix} =
 \begin{pmatrix} \psi_1(\rho) \\ \psi_2(\rho) \exp ({\mbox i}\eta\theta) \end{pmatrix}  \frac{\exp({\mbox i}m\theta)}{\sqrt{2\pi}},
\end{eqnarray}
with two equivalent ground states.  For the first state, 
$m=0$ and the second component has a phase $\exp({\mbox i}\eta \theta )$ relative to the first component. This state is of type  $(0,\eta)$  or $(0,\pm 1)$ for Rashba (upper sign) or Dresselhaus  (lower sign) SO coupling. The second  state has $m=-\eta$  with the first component carrying a phase 
 $\exp(-{\mbox i}\eta \theta )$ relative to the second component.  This state is of type  $(-\eta,0)$  or $(\mp 1,0)$ for Rashba or Dresselhaus SO coupling. 
Of these two types of equivalent  states, in this study we consider only the state 
of type   $(\mp 1,0)$. These states have angular momentum projection $m=0$ and spin $s=1/2$, hence the total angular momentum projection $j_z= m+s_z=\pm 1/2$ and these states are often called    half-quantum vortex states \cite{hpu,hv}.   However, if we consider larger values of $|m| >1$ in Eq. (\ref{12x}), we  can have excited states 
of type $(\mp 2,\mp 1), (\mp 3,\mp 2)$ etc. of larger energy. In an actual nondipolar
 SO-coupled uniform BEC for small $\gamma$  the  infinite ground state of type  $(\mp 1,0)$
becomes a localized  multi-ring soliton.

%The appearance of the stripe state can be understood from the consideration  
 %that  (\ref{spf})  has the following degenerate solutions 
 Equation (\ref{spf}) also  has the following four degenerate solutions
\begin{eqnarray}\label{eq1}
  \begin{pmatrix} \psi_1 \\ \psi_2 \end{pmatrix}=&  
\begin{pmatrix} \cos(\gamma x) \\ -\sin(\gamma x) \end{pmatrix}, \, 
  \begin{pmatrix} \psi_1 \\ \psi_2 \end{pmatrix}=   
\begin{pmatrix} \sin(\gamma x) \\ \cos(\gamma x) \end{pmatrix},
\\
   \begin{pmatrix} \psi_1 \\ \psi_2 \end{pmatrix}=& \begin{pmatrix} \cos(\gamma y) \\ -{\mbox i}\eta \sin(\gamma y) \end{pmatrix},\,    \begin{pmatrix} \psi_1 \\ \psi_2 \end{pmatrix}= \begin{pmatrix} \sin(\gamma y) \\ {\mbox i}\eta \cos(\gamma y) \end{pmatrix},
    \label{eq2}
\end{eqnarray}
with energy eigenvalue
\begin{align}
{\cal E} =-\frac{\gamma^2}{2}. \label{energyan}
\end{align}
The solutions  (\ref{eq1}) and (\ref{eq2}) represent stripes along $x$ and $y$ directions, respectively, with a spatial period of $\pi/\gamma$ in density.     Both sets of solutions have 
uniform density in the sum of the two components: $|\psi_1|^2+|\psi_2|^2 = 1$.
The energy and the density  of the states are the same for both  Rashba and Dresselhaus SO couplings. In an actual nondipolar or dipolar self-attractive and self-repulsive 
 SO-coupled uniform BEC    the  infinite ground state   with stripe 
becomes a localized  stripe  soliton   with the same density and energy for both Rashba and Dresselhaus SO couplings.

A linear combination of the degenerate states (\ref{eq1})  and (\ref{eq2}), e.g.
\begin{align} 
\label{eq3}
  \begin{pmatrix} \psi_1 \\ \psi_2 \end{pmatrix}=\sqrt n   
\begin{pmatrix} \cos(\gamma x) \pm  \cos(\gamma y)\\ -\sin(\gamma x)\mp {\mbox i}\eta \sin(\gamma y  ) \end{pmatrix},
\end{align}
is also a valid degenerate eigenfunction  of Eq. (\ref{spf}) with energy $ {\cal E}= -\gamma^2/2$  and represents an infinite  square-lattice pattern in density of the components as well as in the total density.  In an actual nondipolar self-attractive and self-repulsive 
 SO-coupled uniform BEC,  for medium values of $\gamma$,   the  infinite square-lattice ground state
becomes a localized  square-lattice  soliton   with the same density and energy for both Rashba and Dresselhaus SO couplings.

Equations  (\ref{12x}), (\ref{eq1}) and (\ref{eq2}),  and  (\ref{eq3})   describe   three possible infinite eigenstates of a  uniform binary SO-coupled linear system governed by the Hamiltonian (\ref{abc}). We will see in the following that,   the eigenstates for the localized binary SO-coupled   solitons, controlled by Eq. (\ref{EQ1}),  may  be well approximated by the eigenstates  (\ref{12x}), (\ref{eq1}) and (\ref{eq2}),  and  (\ref{eq3})  times an appropriate Gaussian function, which gives the localization of the soliton.  Hence in the numerical 
calculation employing imaginary-time propagation of the wave function of the  localized binary SO-coupled nondipolar and dipolar solitons,
  we will be using these  approximations  as the initial states.

%For an equal mixture of Rashba and Dresselhaus SO couplings ($i\gamma \partial_x \sigma_y $),  (\ref{spf}) becomes 
%\begin{eqnarray}
%\begin{bmatrix}\label{spf2}
%-\textstyle \frac{1}{2}\nabla^2_{\boldsymbol \rho}&  \gamma \partial_x  \\
%-\gamma\partial_x  & -\textstyle \frac{1}{2}\nabla^2_{\boldsymbol \rho}
%\end{bmatrix}   \begin{pmatrix} \psi_1 \\ \psi_2 \end{pmatrix}
%= {\cal E}   \begin{pmatrix} \psi_1 \\ \psi_2 \end{pmatrix}.
%\end{eqnarray}
%Equation (\ref{spf2}) does not allow $(0,\pm 1)$-type state implied by  (\ref{12x}).
%Equation  (\ref{spf2})  has the spatially-periodic stripe solution (\ref{eq1}).  This solution $(\cos(\gamma x), -\sin(\gamma x))^{\mathrm T}$ with energy ${\cal E} =-\gamma^2/2$ represents stripe in density along $y$ direction.  There is no degenerate solution, similar to (\ref{eq2}),  with the same energy. Thus there can be no solution of type (\ref{eq3})  in this case,  which could represent a square-lattice state with square-lattice pattern in density.    

\section{Numerical Result}
 
 \label{iii}
 
To solve the binary equations   (\ref{EQ1})  numerically, we propagate
these  in time by the split-time-step Crank-Nicolson discretization scheme \cite{bec2009}.
% with {the boundary condition that the wave-function components and their first derivatives vanish at the boundary} while 
As we require to solve the GP equation in the presence of  both SO-coupling and dipolar  interaction,  we  needed  to combine the Open Multiprocessing Programs  for solving the 
dipolar \cite{bec2015,bec2023} and SO-coupled \cite{bec2021} GP equations.
 We employ  space
steps of $dx=dy=0.1$ and a time step of $dt=dx\times dy\times 0.1$ for imaginary-time propagation.   { In all calculations of  stationary states, we  employ  imaginary-time approach {with the 
conservation of total normalization
\begin{equation}
\int_{-\infty}^{\infty} d x \int_{-\infty}^{\infty}dy [n_{1}(x,y)+n_{2}(x.y)] =1
\end{equation}
during time propagation,}
which 
finds the lowest-energy solution of each type. 
 The magnetization 
 \begin{align}
 M= \frac{\int_{-\infty}^{\infty} d x \int_{-\infty}^{\infty}dy [n_{1}(x,y)-n_{2}(x,y)] }{\int_{-\infty}^{\infty} d x\int_{-\infty}^{\infty} dy[n_{1}(x,y)+n_{2}(x,y)]},
 \end{align}
  is
not a good quantum number in the presence of  spin-mixing SO coupling and is left to freely evolve
during time propagation to attain a final converged value
consistent with the parameters of the problem. In  the presence of SO coupling the number of atoms in each component   is not conserved due to the possibility of a spontaneous transfer of atoms from one spin state to the other.
The converged value of the magnetization $M$
was  found to be  negligibly small in most cases for values of $\gamma$ considered in this paper   
indicating an   almost equal number of atoms in the two components.  For $\gamma=0$, the magnetization is strictly equal to zero provided that the initial magnetization is zero.

{  We find different types of soliton with different spatial symmetry properties  in our numerical investigation by imaginary-time propagation.      In order to  obtain the stripe soliton, we use the stripe function   (\ref{eq1}) times  a Gaussian function as the initial state in imaginary-time propagation.  To obtain the square-lattice  soliton in imaginary-time propagation in an efficient way, we use the function (\ref{eq3})  multiplied by a Gaussian function as the initial state.    To  obtain the  circularly-asymmetric   soliton, we  use localized Gaussian functions as the initial state in  imaginary-time  propagation.  Finally, to obtain a 
multi-ring ($\mp 1,0$) soliton, we print the vortex structure, viz. Eq. (\ref{12x}), on the initial Gaussian function. 

 In an actual experiment \cite{14}, by carefully ramping up the intensity of the Raman beams used to create the SO coupling and adjusting the detuning, the system undergoes a  phase transition into the stripe phase. The competition between the interaction energy and the SO-coupling term in the Hamiltonian leads to a ground state where the atoms condense in a superposition of two finite-momentum states, spontaneously breaking the translational symmetry of space. The multi-ring phase can be achieved by imprinting an appropriate  vortex  by a laser on the BEC. 
}

\subsection{ Quasi-2D SO-coupled    binary self-repulsive  nondipolar BEC soliton}
 
\label{a}  

We study the formation of different types of soliton in a quasi-2D SO-coupled   binary self-repulsive and self-attractive dipolar and   nondipolar BEC. For a self-repulsive BEC   the scattering lengths $a$ and $a_{12}$  are positive indicating a repulsive intra-component and inter-component interaction.
Hence, in this system no solitons are possible in the absence of a  dipolar interaction and SO coupling. Both a dipolar interaction  \cite{malomed}  and an SO coupling \cite{quasi-2da} contribute to an attraction in the system and it helps to form a bound soliton.
 Here, we demonstrate that in the presence of SO coupling alone,  the system may  become attractive  and allow the formation of a quasi-2D soliton.

\begin{figure}[!t] 
\centering 
\includegraphics[width= \linewidth]{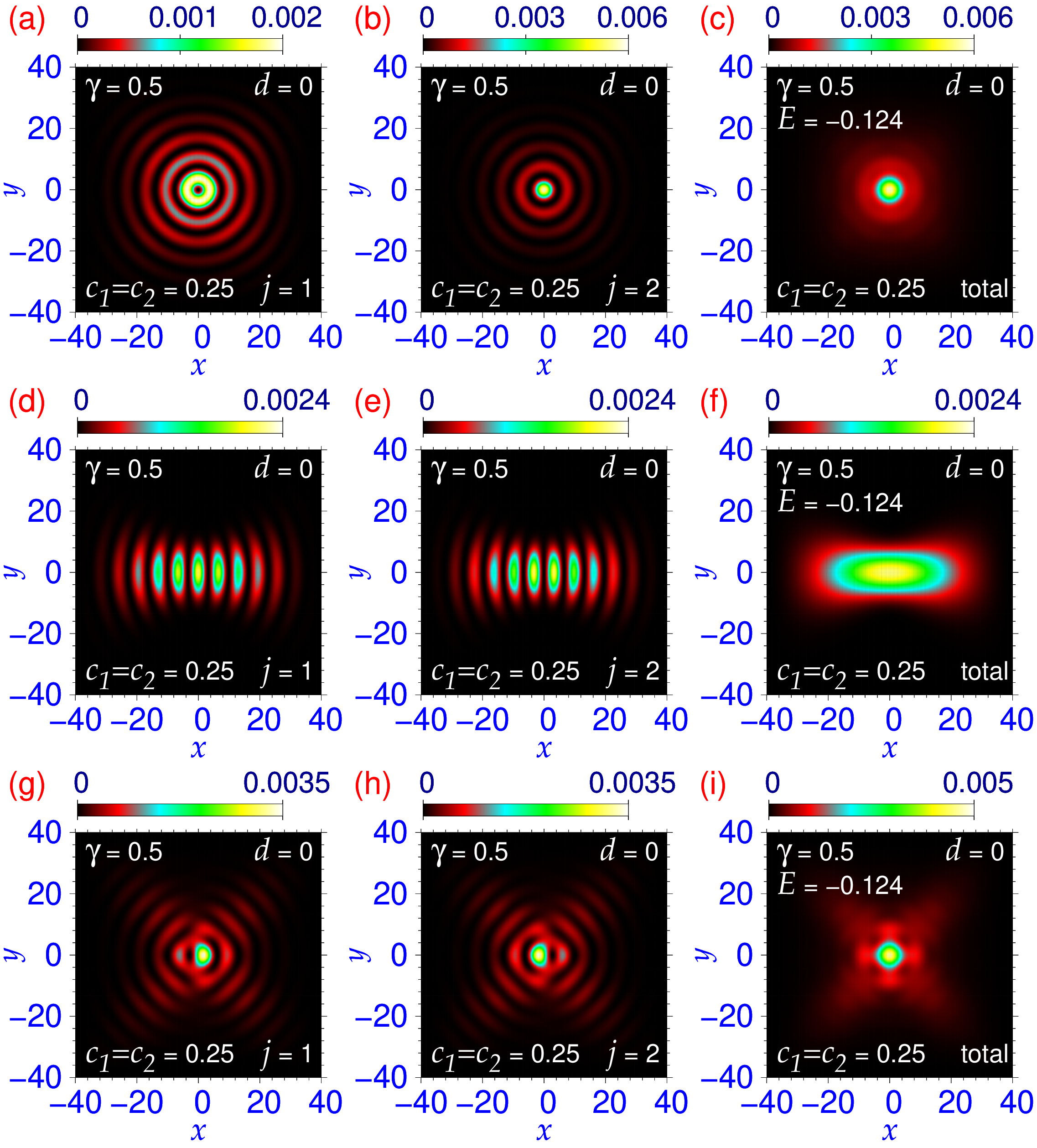}

\caption{ Contour plot of density $n_j$ of a quasi-2D nondipolar  ($d=0$) self-repulsive  multi-ring  $(\mp 1,0)$-type binary    Rashba or  Dresselhaus  SO-coupled BEC soliton 
 for components (a) $j=1$ ($n_1$), (b) $j=2$ ($n_2$) and (c) total density ($n=n_1+n_2$);
 the same of a stripe soliton for components (d) $j=1$ , (e)  $j=2$  and (f) total density; and 
  of a circularly-asymmetric    soliton  
 for  (g) $j=1$, (h) $j=2$ and (i) total density. In this figure and   in Figs. \ref{fig3}, \ref{fig4}, \ref{fig5} and \ref{fig6} the energy is given in the plots of total density in the last column. 
 The parameters are $c_1=c_2=0.25, \gamma =0.5$.  The plotted quantities in all figures of this paper are dimensionless.  }
\label{fig1}

\end{figure}

\begin{figure}[!t] 
\centering 
\includegraphics[width=.95\linewidth]{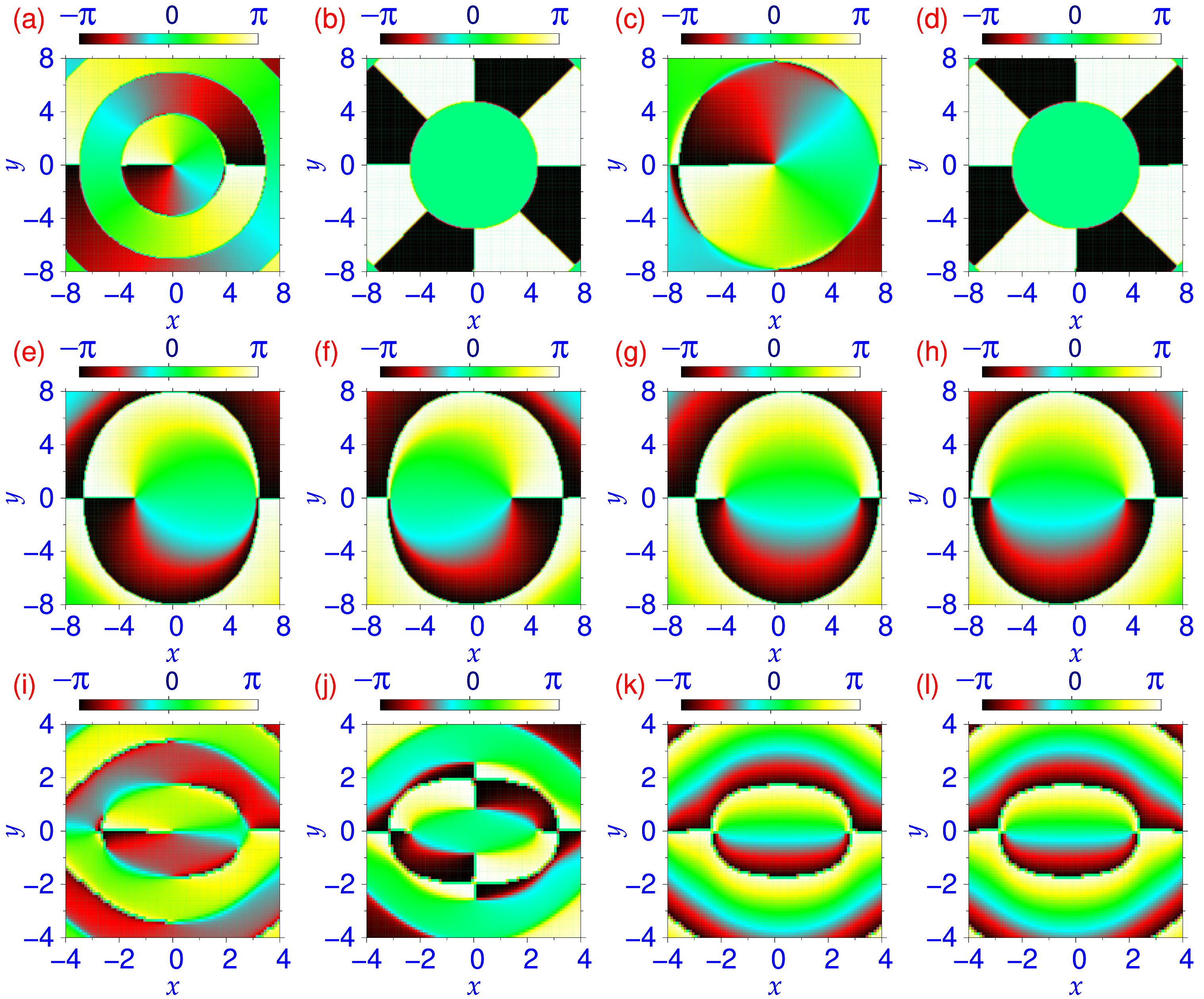}

\caption{ Contour plot of phase of  wave-function components (a) $j=1$, and (b) $j=2$ of the quasi-2D Dresselhaus SO-coupled multi-ring soliton of Figs.  \ref{fig1}(a)-(b);   the same of wave-function   components (c) $j=1$, and (d) $j=2$ of a Rashba  SO-coupled  soliton of Figs.  \ref{fig1}(a)-(b); of wave-function components  (e) $j=1$, and (f) $j=2$ of the Dresselhaus SO-coupled  soliton of Figs.  \ref{fig1}(g)-(h); of wave-function components  (g) $j=1$,  (h) $j=2$ of 
the Dresselhaus SO-coupled dipolar soliton of Figs.  \ref{fig4}(g)-(h);
of wave-function components  (i) $j=1$,  (j) $j=2$ of 
the Dresselhaus SO-coupled dipolar soliton of Figs.  \ref{fig5}(a)-(b);
of wave-function components  (k) $j=1$,  (l) $j=2$ of 
the Dresselhaus SO-coupled dipolar soliton of Figs.  \ref{fig5}(d)-(e).}
\label{fig2}

\end{figure}

   Without losing generality we   consider  a nondipolar BEC with  a small value of the nonlinearity parameters  $c_1=c_2=0.25$ in Eqs. (\ref{EQ1}), so that a soliton can be formed for a small value of the SO-coupling strength $\gamma$, which facilitates both theoretical and numerical investigations. For a small    $\gamma$ ($\gamma = 0.5$) we  find three types of degenerate solitons in a self-repulsive nondipolar BEC. 
First,  there is   a $(\mp 1,0)$-type multi-ring soliton   for Rashba  (upper sign) or Dresselhaus (lower sign) SO coupling.
In Fig. \ref{fig1} we display the contour plot of   density  of the central region 
of components (a) $j=1$, (b) $j=2$, and (c) total density   of a  $(\mp 1,0)$-type  multi-ring soliton.  The left (middle) column shows the density of component $j=1$ ($j=2$), whereas the right column illustrates the total density.  The density is independent of whether we employ Dresselhaus SO coupling or Rashba SO coupling. However, the phase of this (+1,0)-type  Dresselhaus SO-coupled  soliton is distinct from a $(-1,0)$-type Rashba SO-coupled soliton.
  In Figs. \ref{fig2}(a)-(b) we display the contour plot of the phase of wave function  components $j=1,2$ of the   $(+1,0)$-type Dresselhaus SO-coupled 
soliton of Figs. \ref{fig1}(a)-(c).  In Fig.   \ref{fig2}(a) there is a  phase drop of 
$+ 2\pi$   under a complete rotation, 
indicating an angular momentum projection of $+1$ (vortex of unit angular momentum) in component $j=1$ and 0 in component $j=2$.  
The corresponding phases for Rashba SO coupling are illustrated in 
Figs. \ref{fig2}(c)-(d) showing the phases of the wave-function components $j= 1,2$ of the SO-coupled soliton. In this case in Fig.  \ref{fig2}(c) we find a phase drop of  $- 2 \pi$  under a complete rotation around the origin, indicating an angular momentum projection of $-1$   (anti-vortex of unit angular momentum) in component $j=1$ and 0 in component $j=2$, viz. Fig. \ref{fig2}(d).
  In addition to the multi-ring solitons, for $\gamma=0.5$ we  can also have
  stripe solitons with  stripes in density, of a spatial period $\pi/\gamma$, of both the components as shown in Figs. \ref{fig1}(d)-(f) for components $j=1,2$ and total density, respectively. 
  However, the total density of the two components is  without any stripe pattern as shown in Fig. \ref{fig1}(f).  Moreover, we have a { circularly-asymmetric} soliton as  shown in Figs.  \ref{fig1}(g)-(i), where  we display the density of the two components and the total  density, respectively. 
A close examination of the soliton components $j=1,2$ in Figs. \ref{fig1}(g)-(h), respectively, reveals the presence of asymmetrically located vortex-anti-vortex pairs, explicitly  demonstrated through a contour plot of the phase of these wave-function components in Figs. \ref{fig2}(e)-(f).   % In Figs. \ref{fig2}(g)-(h) we display the central part of the density of these two components.
  The component densities are zero (black)  in Figs. \ref{fig1}(g)-(h)  at the position of vortex and anti-vortex in Figs. \ref{fig2}(e)-(f), but the total density is not zero at these points,  indicating that these are coreless. 
  These three types  of solitons are degenerate with the numerical  energy ${\cal E}=-0.124$    close to  its theoretical estimate  ${\cal E}= \gamma^2/2= -0.125$ for $\gamma=0.5$. 
  In this case no square-lattice soliton could be found. The scenario of soliton formation does not change for  a self-attractive nondipolar BEC, where there are also these three types of soliton (result  not presented in this paper).

\begin{figure}[!t] 
\centering
\includegraphics[width=\linewidth]{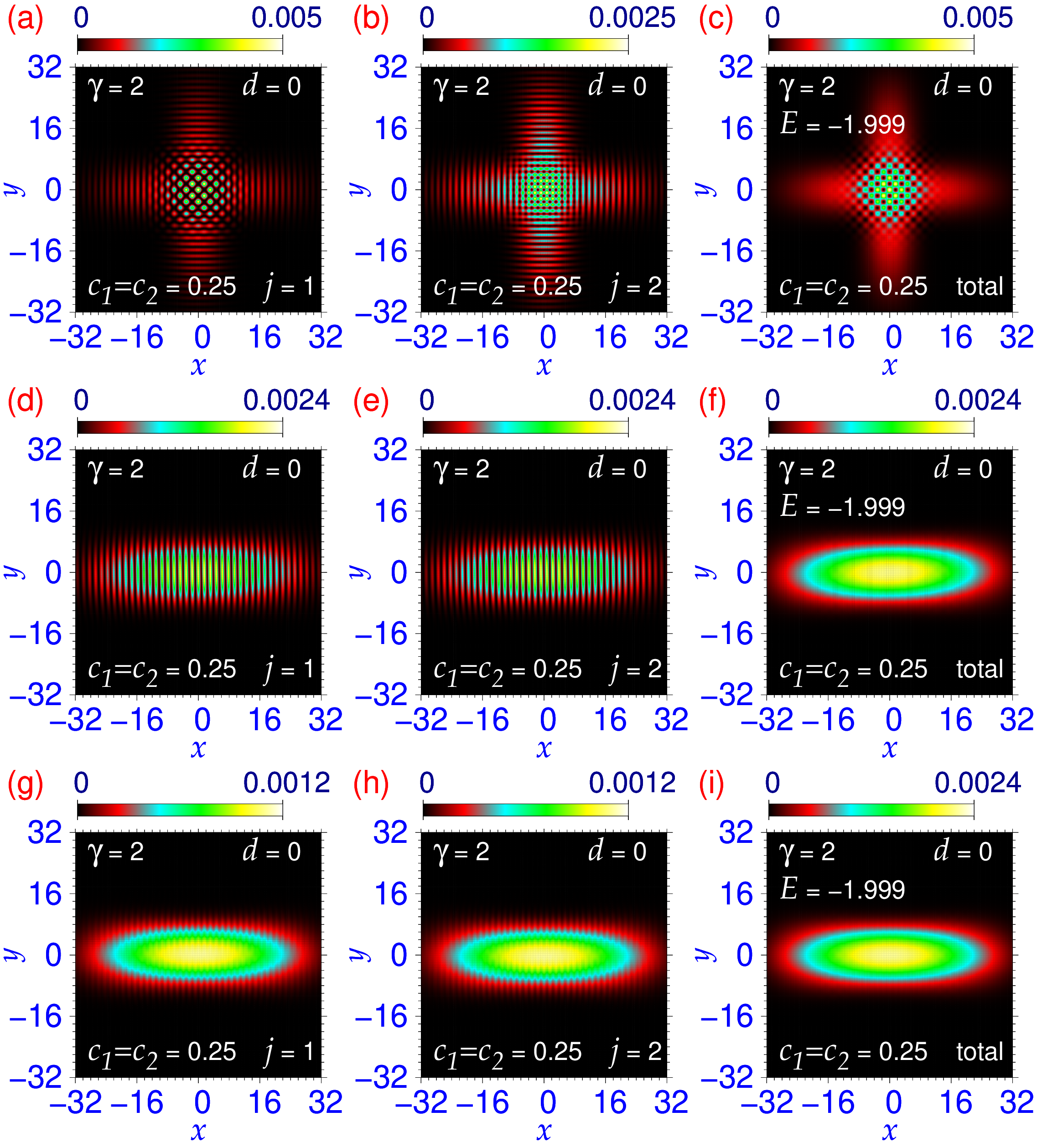}

\caption{Contour plot of dimensionless   density 
$n_{2D}(x,y)$ of   (a)-(c) a square-lattice soliton, (d)-(f) a stripe soliton, and 
(g)-(i) a circularly-asymmetric soliton in a self-repulsive nondipolar SO-coupled BEC. 
The left [middle] column shows the density of component $j=1 (n_1)$ [$j=2 (n_2)$], whereas the right column depicts  the total density. The parameters of the model are $c_1=c_2=0.25, d=0$ and $\gamma=2$.
}
\label{fig3}

\end{figure}

As $\gamma$ is increased,  a new scenario of soliton formation emerges.  However, the types of solitons for $\gamma =1$ to $\gamma =3$    are quite similar and here we concentrate on $\gamma=2$.
For $\gamma =2$, we find three types of degenerate solitons in a quasi-2D nondipolar self-repulsive SO-coupled  BEC.  
The    $(\mp 1,0)$-type
multi-ring soliton, viz. Figs. \ref{fig1}(a)-(c),
ceases to exist and gives rise to a new type of soliton: a square-lattice soliton with square-lattice modulation in density.  
This soliton is illustrated in Figs. \ref{fig3}(a)-(c)  through a contour plot of the density of (a)  component $j=1$,  (b) component $j=2$,  and (c) total density. The stripe soliton and the circularly-asymmetric soliton continue to exist.
The resultant densities of the stripe soliton, with a spatial period  of $\pi/\gamma$, are  illustrated in Figs. \ref{fig3}(d)-(f). The densities of a circularly-asymmetric soliton %and a four-wing soliton 
are depicted in Figs. \ref{fig3}(g)-(i), which are  very different from the same of the {circularly-asymmetric}  soliton of Figs.  \ref{fig1}(g)-(i) for $\gamma=0.5$. Also, there is no prominent vortex-anti-vortex structure in the wave functions of the two components as in the {circularly-asymmetric} soliton shown in Figs.
 \ref{fig1}(g)-(h)     for $\gamma=0.5$,  viz. Fig. \ref{fig2}(e)-(f).
%   and \ref{fig3}(j)-(l).  
%Although, the solitons of Fig. \ref{fig1} for $\gamma =0.5$ evolve to the solitons of Figs. \ref{fig3}(a)-(i), visually they are different.  
The stripe solitons of Figs. \ref{fig1}(d)-(f) and \ref{fig3}(d)-(f) are quite similar. However,   in Figs.  \ref{fig3}(a)-(f)  the spatially periodic pattern  has a small period consistent with $\gamma =2$.  Although, there is a stripe pattern in density for both $\gamma=0.5$ and $\gamma=2$,
the positions of maxima in component $j=1$ coincide with the minima in component $j=2$, thus resulting in a total density without modulation.  For $\gamma =2$, the total density of the stripe soliton in Fig. \ref{fig3}(f) is identical to the same of the {circularly-asymmetric} soliton in Fig. \ref{fig3}(i), although the component densities are distinct. 
 Hence, for a description of total density and energy, these two binary solitons behave like a unified whole, possessing the same total density and energy.  We will see in the following that even in the presence of dipolar interaction the total densities of the stripe soliton and the  {circularly-asymmetric} soliton continue to be equal. In this nondipolar case there is an $x\leftrightarrow y$ symmetry and there is a stripe and {a circularly-asymmetric} solution aligned along the $y$ axis with the same density and energy as the solitons in Figs. \ref{fig3}(d)-(f) and Figs. \ref{fig3}(g)-(i), respectively (not shown here). {The solitons in Figs. \ref{fig3}(d)-(f) and Figs. \ref{fig3}(g)-(i), although are limited in the identical spatial region, they are distinct.  The former has stripes (dark regions) along the $y$ axis; the area of the dark and the bright regions are the same.  In the latter, there is no such dark region and the same number of atoms are distributed over the whole area. Consequently, the density in this case is half of the stripe soliton, compare the maximum densities 0.0024 and 0.0012 in the two cases.}     
 The numerical energy of  the degenerate solitons illustrated in Fig. \ref{fig3} is ${\cal E}=-1.999$ in perfect agreement with the theoretical estimate ${\cal E}=-\gamma^2 /2$, which for $\gamma=2$ is ${\cal E}=-2$.
  {The degeneracy 
  of the  states depicted in Fig. \ref{fig3},
  for small values of nonlinearities ($c_1=0.25, c_2=0.25$ and $d=0$), is a consequence  of the degeneracy of the states in the
  linear problem in  Sec. \ref{ii}, viz. Eqs. \ref{12x}, (\ref{eq1})-(\ref{eq3}). The degeneracy will disappear for a larger value of the nonlinearities.}
     For a self-attractive nondipolar  SO-coupled BEC,  we have the same three types of solitons (result not elaborated here).

 This spatially-periodic pattern in two and one dimension of the square-lattice and stripe solitons of Figs. \ref{fig3}(a)-(c) and Figs. \ref{fig3}(d)-(f), respectively,  
is consistent with the periodic supersolid pattern in two \cite{adhisol} and one \cite{baym1,baym2} dimensions.  The formation of such supersolid pattern in an SO-coupled spinor BEC has long been demonstrated \cite{14,2020} and studied \cite{stripe}. The distribution of matter on a  square lattice is prominent in  Figs. \ref{fig3}(a)-(c), not only in the component densities but also 
in the total density.
The present square-lattice soliton  is a consequence of the SO coupling and breaks { continuous}  translational symmetry as required in a supersolid \cite{s1,s2,s3}.  
The {circularly-asymmetric} soliton of Figs. \ref{fig3}(g)-(i)  is also distinct.

\begin{figure}[!t]
\centering
\includegraphics[width=\linewidth]{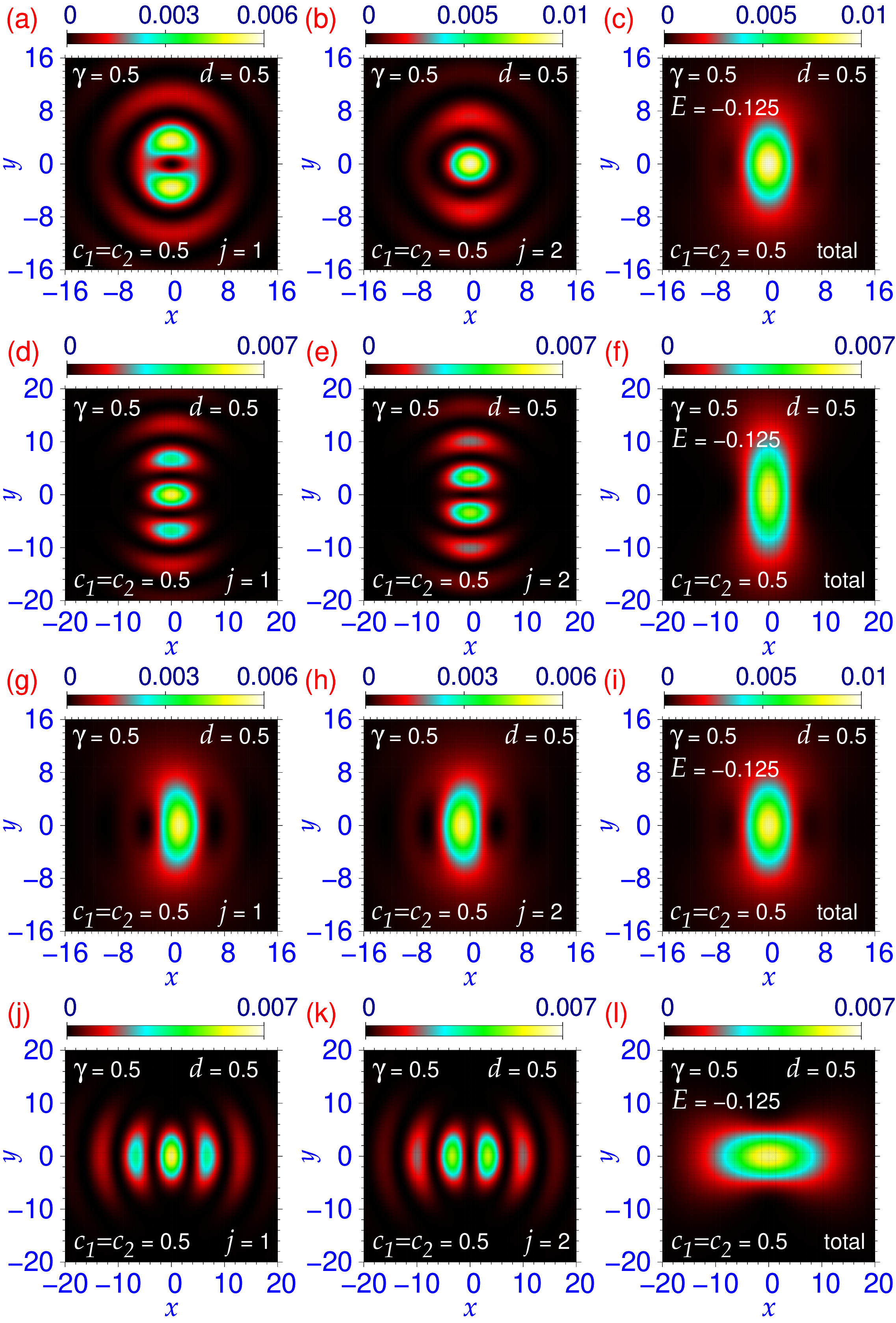}

\caption{Contour plot of dimensionless   density 
$n_{2D}(x,y)$ of  (a)-(c) a multi-ring soliton, 
 (d)-(f)  a stripe soliton with stripes along the   $y$ axis,
(g)-(i) a  circularly-asymmetric soliton, and (j)-(l)  a stripe soliton with stripes along the $x$ axis.
The left [middle] column shows the density of component $j=1 (n_1)$ [$j=2 (n_2)$], whereas the right column depicts  the total density. The parameters of the model are $c_1=c_2=d=0.5,$ and $\gamma=0.5$.}

\label{fig4}

\end{figure}

\subsection{ Quasi-2D SO-coupled  binary self-repulsive and self-attractive  dipolar BEC soliton}
 
\label{b}

 The scenario of soliton formation dramatically changes if we consider dipolar atoms with the polarization direction lying along the $y$ axis in the quasi-2D plane. This  dipolar interaction is strongly anisotropic in the quasi-2D $x$-$y$ plane and it  will  break the circular symmetry of the system.
 With the introduction of the dipolar interaction, the SO-coupled self-repulsive dipolar system acquires additional attraction and the solitons  become more compact and of smaller spatial dimension compared to the nondipolar case considered in  Sec. \ref{a}.   For $\gamma=0.5$, the multi-ring, stripe and the {circularly-asymmetric} solitons continue to exist in this case, with much smaller size,  as shown in Figs. \ref{fig4}(a)-(l),  for $c_1=c_2=d=0.5$. Although, it is possible to have a quasi-2D dipolar soliton with the polarization direction in the quasi-2D plane \cite{malomed}, there is no such soliton for these sets of parameters without an SO-coupling interaction.  The dipolar nonlinearity $d=0.5$ is too weak to support such a soliton.   
The multi-ring soliton presented in Figs. \ref{fig4}(a)-(c)    is of the ($\mp 1,0)$ type and has the phase  structure similar to the same illustrated in Figs. \ref{fig2}(a)-(d) (not explicitly shown in this paper). The stripe  soliton presented in Figs. \ref{fig4}(d)-(f)  has a spatially periodic stripe along the $y$ axis.
   As  in the nondipolar case, the two components of the spatially asymmetric soliton presented in Figs.  \ref{fig4}(g)-(h) host a vortex-anti-vortex structure,
 which is explicit in the plot of the phase of these wave functions in Figs. \ref{fig2}(g)-(h). The total density and energy of the multi-ring soliton presented in Figs. \ref{fig4}(a)-(c)  are  the same as the  {circularly-asymmetric} soliton illustrated in  Figs. \ref{fig4}(g)-(i), although the component densities are distinct.  Finally, we find a stripe soliton with stripe along  $x$ axis shown in Figs. \ref{fig4}(j)-(l), which has practically the same density and also the energy as the stripe soliton with stripe along the $y$ axis, indicating an approximate $x \leftrightarrow y$ symmetry for a small $d=0.5$.   All solitons  of Fig. \ref{fig4} are   degenerate with energy ${\cal E} = -0.125$.  
   We could not find any other type of soliton in this case.

\begin{figure}[!t]
\centering
\includegraphics[width=\linewidth]{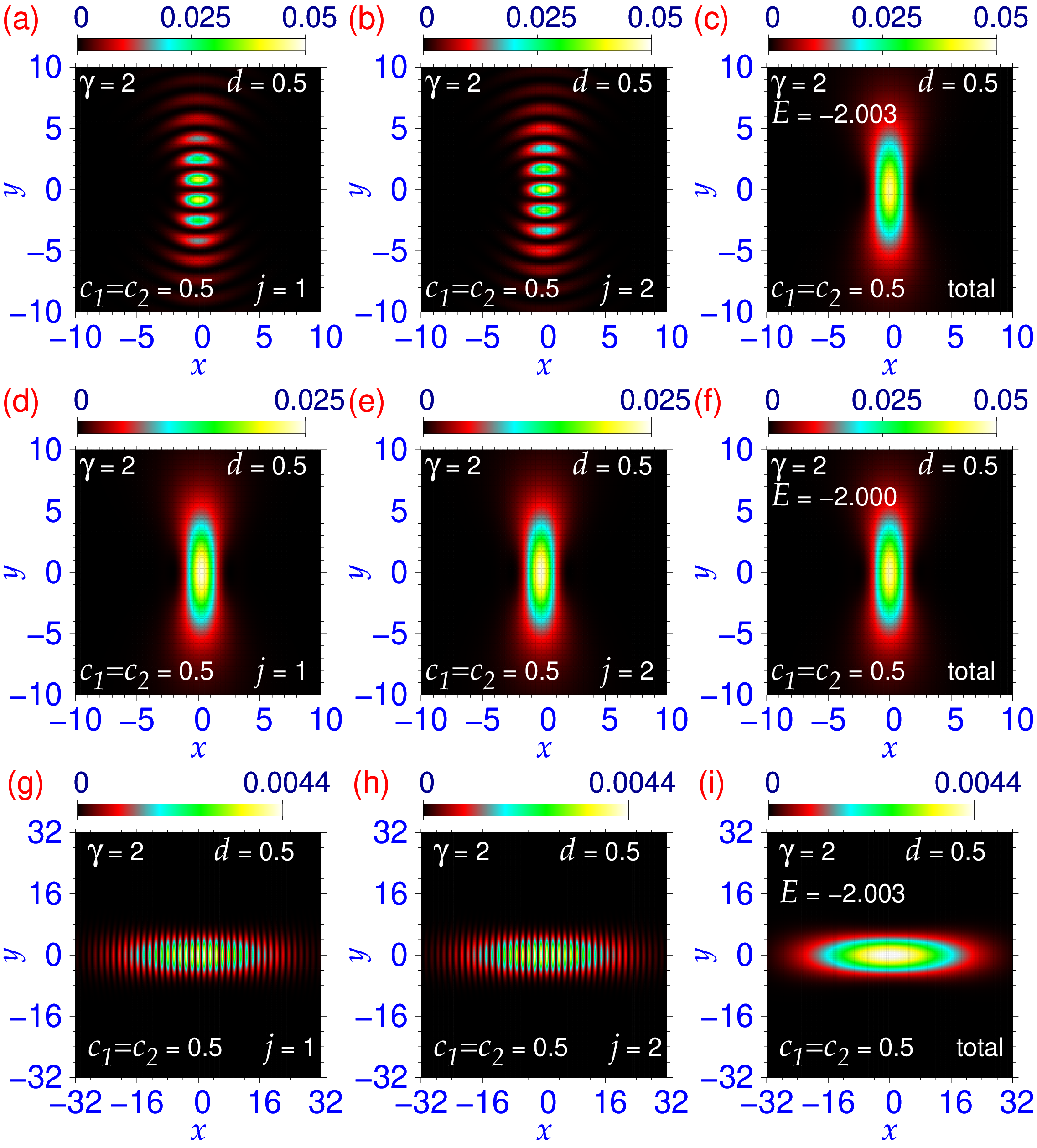}

\caption{Contour plot of dimensionless   density 
$n_{2D}(x,y)$ of   (a)-(c)  a stripe soliton with stripes along the polarization  $y$ axis,
(d)-(f) a {circularly-asymmetric} soliton, and (g)-(i) a stripe soliton with stripes along the  $x$ axis.
The left [middle] column shows the density of component $j=1 (n_1)$ [$j=2 (n_2)$], whereas the right column depicts  the total density. The parameters of the model are $c_1=c_2=d=0.5, d=0.5$ and $\gamma=2$.}

\label{fig5}

\end{figure}

As the SO-coupling strength $\gamma$ increases beyond $\gamma=1$, there is some change in the scenario of the formation of a quasi-2D  soliton in a binary self-repulsive dipolar BEC,  when compared to the same for $\gamma=0.5$ illustrated in Fig. \ref{fig4}. 
Due to the combined attraction  of dipolar interaction and SO coupling,  for $\gamma=2, c_1=c_2=d=0.5$, 
the  multi-ring $(\mp 1,0)$-type soliton is squeezed on to the polarization $y$ axis and evolves into    a  stripe soliton with stripe along the $y$ axes as shown in  
Figs. \ref{fig5}(a)-(c), obtained in imaginary-time propagation employing a Gaussian  initial state imprinted with $(\mp 1,0)$ angular momenta structure. The converged stripe soliton  maintains the approximate  $(\mp 1,0)$ angular momenta structure of the two components. 
The phase of the wave functions of the  two components is presented in Figs. \ref{fig2}(i)-(j).  In Fig. \ref{fig2}(i) we find a phase drop of $2\pi$ under a complete rotation around the origin corresponding to a an angular momentum projection of $+1$ (vortex of unit angular momentum) in the first component. There is no such vortex in the second component, viz. Fig. \ref{fig2}(j). In addition, for these parameters, we find 
   a circularly-asymmetric soliton as presented in   Figs. \ref{fig5}(d)-(f),
and a stripe soliton with stripe along the $x$ axes as illustrated in  Figs. \ref{fig5}(g)-(i). 
The    circularly-asymmetric soliton  possesses  a  vortex-anti-vortex structure as shown in Fig. \ref{fig2}(k)-(l).
The square-lattice soliton is not possible in this case;  due to the dipolar attraction along the polarization $y$ axis this soliton is squeezed on the $y$ axis and eventually destroyed. 
 The stripe solitons presented in Figs. \ref{fig5}(a)-(c) and Figs. \ref{fig5}(g)-(i), respectively, are degenerate ground states with energy $E=-2.003$, more compact and of smaller size  compared to the same solitons in Figs. \ref{fig4}(d)-(f) and in Figs. \ref{fig4}(j)-(l) for $\gamma =0.5$.  The added attraction due to the increased SO-coupling strength  is responsible for this.  Again,  the total density and energy of these two solitons in Figs. \ref{fig5}(a)-(f) are almost identical, although the component densities are distinct, indicating a universal behavior.   Of these three solitons presented in Fig. \ref{fig5} for $\gamma=2$, the {circularly-asymmetric} soliton  is an excited state with  energy ($E=-2.000$).
In the presence of strongly anisotropic dipolar interaction in the quasi-2D plane, there is no $x \leftrightarrow y$ symmetry and the stripe soliton aligned along the $x$ axis, viz. Figs. \ref{fig5}(a)-(c), has different  density structure from the one aligned along the $y$ axis, viz.  Figs. \ref{fig5}(g)-(i).

\begin{figure}[!t] 
\centering
\includegraphics[width=\linewidth]{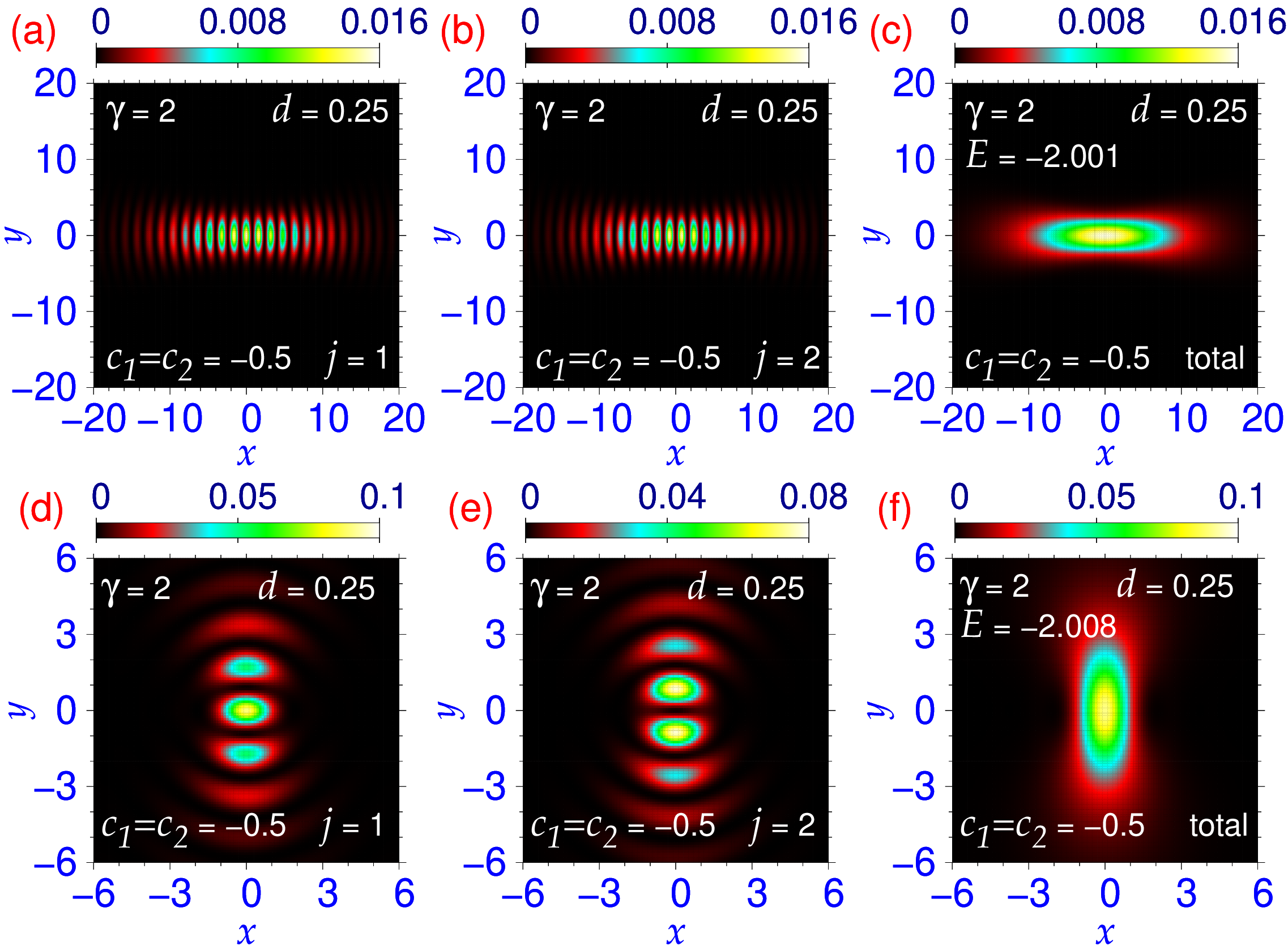}  

\caption{Contour plot of dimensionless   density 
$n_{2D}(x,y)$ of (a)-(c)  a stripe soliton with stripes along the   $x$ axis, and 
  (d)-(f)  a stripe soliton with stripes along the polarization  $y$ axis.
The left [middle] column illustrates the density of component $j=1 (n_1)$ [$j=2 (n_2)$], whereas the right column depicts  the total density. The parameters of the model are $c_1=c_2=-0.5, d=0.25$ and $\gamma=2$.}

\label{fig6}

\end{figure}

Although, the principal interest of  this paper is the formation  of quasi-2D solitons  in a self-repulsive SO-coupled dipolar BEC, it is also appropriate to investigate, if any, the changes in the structure of   such solitons  as  the system changes from self-repulsive to self-attractive. To this end, in Fig. \ref{fig6} we illustrate the densities of the quasi-2D solitons for a self-attractive SO-coupled dipolar BEC for $\gamma=2, c_1=c_2=-0.5, d=0.25.$ 
In this case, we  reversed the sign of the strengths $c_1$ and $c_2$  of the contact interaction and reduced the same of the dipolar interaction $d$ from 0.5  to 0.25. 
  A larger value of $d$ will imply more attraction, which will make the solitons to shrink to have a much smaller size.  However, for this set of parameters ($c_1=c_2=-0.5, d=0.25$)  there  is no soliton in the absence of SO coupling;  the dipolar nonlinearity ($d=0.25$) is too small to support a stable soliton.  
The circularly-asymmetric soliton of Figs. \ref{fig5}(d)-(f) is not possible in this case due to the added contact attraction.  This soliton is eventually destroyed and turns to a stripe soliton with stripe along the $y$ axis. 
In this case, we find  only two types of solitons: a stripe soliton   with stripe along the $x$ axis with energy $E=-2.001$ shown in Figs. \ref{fig6}(a)-(c) and  a stripe soliton with stripe along the $y$ axis with energy $E=-2.008$  depicted  in Figs. 
\ref{fig6}(d)-(f).  
Due to an increased  contact attraction,  the solitons  of Fig. \ref{fig6} are more compact and of much smaller size, when compared to the same in Fig. \ref{fig5}. 
 In the presence of a stronger attraction,  away from the linear limit, there is no degeneracy of the solitons illustrated in Fig. \ref{fig6}.

\section{Summary and Discussion}
 \label{iv}
 
To search for a quasi-2D soliton  in a uniform  binary   SO-coupled dipolar and nondipolar, self-repulsive  BEC, using a numerical solution of a mean-field model,  we identify  a multi-ring soliton, a circularly-asymmetric soliton, a spatially-periodic stripe soliton,  and a square-lattice soliton for  small to medium  strengths of SO coupling $\gamma$  ($\gamma \lessapprox 4$).  We consider Rashba and Dresselhaus SO couplings in this paper.
In all cases the density obtained with Rashba and Dresselhaus SO couplings are the same although the wave functions and the phases of the solitons are different, viz. Fig. \ref{fig2}.
For  a small $\gamma$ ($\gamma \lessapprox 1$), 
there are three types of degenerate states in a nondipolar  self-repulsive SO-coupled BEC, viz. Fig. \ref{fig1},: a  
$(\pm 1,0)$-type multi-ring soliton, a circularly-asymmetric soliton, and a stripe soliton. 
A $(\pm 1,0)$-type  multi-ring soliton has a vortex or anti-vortex   of unit vorticity  in one component and no vortex in the other component in addition to  having  a multi-ring modulation in density of the components without  such a  modulation in total density, viz. Fig. \ref{fig1}(a)-(c). A stripe soliton has a spatially-periodic stripe in the density of both components and a circularly-asymmetric Gaussian-type  total density without stripe,  viz., for example,  Fig. \ref{fig1}(d)-(f).   A circularly-asymmetric soliton of  Figs. \ref{fig1}(g)-(i)
      has a vortex and an anti-vortex in the  wave function  of each component,  viz.,  
      Figs. \ref{fig2}(e)-(f).
 For a medium $\gamma$ ($1 \lessapprox \gamma \lessapprox 4$) in a nondipolar self-repulsive SO-coupled BEC  we find three types of degenerate  solitons, viz. Fig. \ref{fig3}: (a)-(c) a spatially-periodic square-lattice soliton, (d)-(f) a  stripe soliton and a  (g)-(i){ circularly-asymmetric} soliton with vortex-anti-vortex structure.   
 
  For a dipolar self-repulsive SO-coupled BEC, for a small SO coupling $\gamma =0.5,  $
we find four types of degenerate solitons: a multi-ring soliton, a stripe soliton with stripe along the $x$ or $y$ axis, and a circularly-asymmetric soliton, as illustrated in Fig. \ref{fig4}.  In the same system, for a large $\gamma (=2)$, we have  three types of solitons: a stripe soliton with stripe along the $x$ or $y$ axis,  and a circularly-asymmetric soliton aligned along the $y$ axis, viz. Fig. \ref{fig5}. The stripe solitons with the stripe along the $x$ axis and the  polarization $y$ axis are degenerate with the circularly-asymmetric soliton emerging as an excited state.
 In a weakly self-attractive dipolar  SO-coupled BEC, we find only two types of solitons. e.g.  a stripe soliton with stripe along the $x$ axis or the $y$ axis.  Of these two type the one with stripe along the $y$ axis is the ground state, viz. Fig. \ref{fig6}. 

The spatially-periodic density in stripe \cite{14,baym1,2020,stripe} and square-lattice solitons is a  possible  manifestation of supersolid-like \cite{s1,s2,s3,s4} properties. For the small nonlinearities considered in this paper,  different types of possible solitons in a self-repulsive SO-coupled nondipolar BEC  for a  fixed $\gamma$ are all  degenerate  with an energy very close to the theoretical estimate ${\cal E}=-\gamma^2 /2$.   For a small $\gamma$, the solitonic states in a uniform SO-coupled BEC have a  very large spatial extension.
For a self-repulsive SO-coupled dipolar BEC, for a small $\gamma=0.5$ all types of solitons are degenerate; for a medium $\gamma =2$  this degeneracy is partially broken and the stripe soliton with stripe along the $x$ or $y$ axis are degenerate   ground states, whereas the circularly-asymmetric soliton is an excited state.
For a medium $\gamma$,  the total density of the stripe  and {circularly-asymmetric} solitons are  found to be almost equal, which seems to be a universal behavior.
  All these states are dynamically stable, as we verified by real-time simulation over a long period of time     (result not presented in this paper).  
  The stripe,  square-lattice, and circularly-asymmetric  solitons  were also found in an  SO-coupled  nondipolar  quasi-2D spin-one  \cite{adhisol} and spin-two \cite{pardeep}  BECs. However, it is more difficult to perform an experiment in an SO-coupled  spin-one or spin-two BEC with three or five  spin components compared to an SO-coupled binary BEC with two spin components. Hence we believe that the present study could motivate experiments in  the search of square-lattice states in an SO-coupled    binary BEC.
  The quasi-1D stripe state in a trapped SO-coupled     spinor BEC has already been observed and investigated \cite{14}.
  The square-lattice soliton is dynamically robust and deserve   further  theoretical and experimental  investigation.

%\end{document}

\section*{ACKNOWLEDGMENT}

The author  
 acknowledges support by the Conselho Nacional de Desenvolvimento  Científico e Technológico  (Brazil) grant  303885/2024-6.

 \section*{DATA AVAILABILITY}
The data that support the findings of this article are not
publicly available upon publication because it is not technically feasible and/or the cost of preparing, depositing, and
hosting the data would be prohibitive within the terms of this
research project. The data are available from the authors upon
reasonable request.

\end{document}